\documentclass[reprint,aps,pre,showkeys,showpacs,superscriptaddress,onecolumn,amsmath,amssymb,longbibliography]
{revtex4-1} 
\usepackage{graphics}
\usepackage{bm}
\usepackage{subfigure}
\usepackage{epsfig}
\usepackage[bookmarks,colorlinks,citecolor=blue,linkcolor=blue]{hyperref}
\usepackage{natbib}
\usepackage{siunitx}
\usepackage{soul}
\usepackage{array}
\usepackage{multirow}

\DeclareSIUnit\Molar{\textsc{m}}

\usepackage{xcolor}
\usepackage[normalem]{ulem}

\usepackage{graphicx}	
\begin{document}
\title{Frictional Contact Network in Dense Suspension Flow}

\author{Shweta Sharma}
\affiliation{Department of Macromolecular Science and Engineering, Case Western Reserve University, Cleveland, OH, 10040, USA}
\author{Abhishek Sharma}
\affiliation{Department of Chemical Engineering, Albert Nerken School of Engineering,
The Cooper Union for the Advancement of Science and Art, New York, NY, 10003, USA}
\author{Abhinendra Singh}
\date{March 2025}
\email{abhinendra.singh@case.edu}
\affiliation{Department of Macromolecular Science and Engineering, Case Western Reserve University, Cleveland, OH, 10040, USA}

\begin{abstract}
Dense particulate suspensions often exhibit a dramatic increase in viscosity in response to external deformation. This shear thickening behavior has been related to a transition from lubricated, unconstrained pairwise motion to a frictional contact network (FCN) at high stresses. Here, we study the characteristics of the FCN formed during shear thickening to investigate the role of constraints, emphasizing the impact of resistance to gear-like rolling. We contrast the FCN formed by sliding friction alone with that formed by particles with sliding and rolling constraints.
Particles with sliding constraints only form a highly interconnected network with primary force chains in the compressive direction, which requires orthogonal support from other force chains.
However, orthogonal support is not required for mechanical stability when particles have both sliding and rolling constraints. 
In addition, the force chains appear linear and longer, reducing the jamming volume fraction for rough/faceted particles.
Finally, we propose a novel mechanical stability picture for rough/faceted particles with sliding and rolling constraints, which is crucial for understanding the flow behavior of real-life suspensions.

\end{abstract}

\maketitle

 
\textit{Introduction-} Considering a dense suspension of non-Brownian hard particles ($R\gtrsim $ \SI{1}{\micro\meter}) dispersed a density matched viscous Newtonian liquid. Given the particle size and limit of the hard sphere, the suspension has no intrinsic time or stress scale. Dimensional analysis would suggest that viscosity $\eta$ is independent of stress $\sigma$ or shear rate $\dot{\gamma}$~\cite{Morris_2020, Denn_2014, Cates_2014}. These seemingly simple systems exhibit complex behaviors, where macroscopic phenomena such as yielding, shear thinning, shear thickening, and shear jamming (SJ) occur~\cite{Mari_2014,Morris_2020,Singh_2018,guazzelli_2018, Singh_2019, Jamali_2020}. These bulk responses have been shown to arise from intricate particle interactions, including friction, repulsion, attraction, and hydrodynamics \cite{Mari_2014,Morris_2020,Singh_2018, Singh_2019, Singh_2020}. 

Mean-field approaches model shear thickening by attributing viscosity increases to a stress-activated crossover from unconstrained to constrained tangential pairwise motion above an ``onset stress'' \cite{Wyart_2014,Singh_2018,Morris_2020,Ness_2016,Singh_2022, Guy_2015, Guy_2018}. At low stresses, the particles are in a lubricated state, and the viscosity diverges at $\phi_J^0$. In contrast, at high stresses, frictional contacts dominate, leading to a viscosity divergence at a lower volume fraction $\phi_J^{\mu}$. The Wyart-Cates (WC) model, a key framework, uses a two-state approach with a scalar order parameter, the fraction of frictional contacts, to interpolate between these states with an increase in stress, which mainly leads to three forms of flow curves~\cite{Wyart_2014}. At volume fractions $\phi \ll \phi_J^\mu$, viscosity increases continuously with shear rate, leading to continuous shear thickening (CST). At large enough volume fractions $\phi_c<\phi<\phi_J^\mu$, the flow curve $\eta(\dot{\gamma})$ becomes non-monotonic and S-shaped, and the thickening becomes discontinuous (DST). Finally, for $\phi>\phi_J^\mu$, the frictional branch is jammed\textit{, i.e.}, the suspension flows at low stresses but is in a shear-jammed state (SJ) at large stresses.
 Although traditionally focused on sliding friction, recent studies suggest that rolling friction should be considered especially to model shear thickening behavior with real-world suspensions composed of rough particles (spherical particles with asperities or faceted particles) or with adhesive interactions \cite{Singh_2020,Lootens_2005,Hsiao_2017, Pradeep_2021,d2023role}. Roughness at the particle level~\cite{Hsu_2018,hsiao2019experimental,Hsiao_2017, Lootens_2005} and interfacial chemistry~\cite{James_2018, James_2019} are known to enhance shear-thickening. Singh et al.~\cite{Singh_2020, Singh_2022, Singh_2023} demonstrated that this enhanced shear thickening can be understood in terms of enhanced ``effective friction'' originating mainly from rolling friction due to asperities coming into contact or adhesive interactions between particles. Recent experimental studies using rough/faceted particles show the crucial importance of rolling friction, making it an important candidate for a comprehensive understanding of real-world sheared suspensions~\cite{scherrer2024sliding,d2023role}.

\begin{figure}
    \centering
    \includegraphics[width=0.7\linewidth]{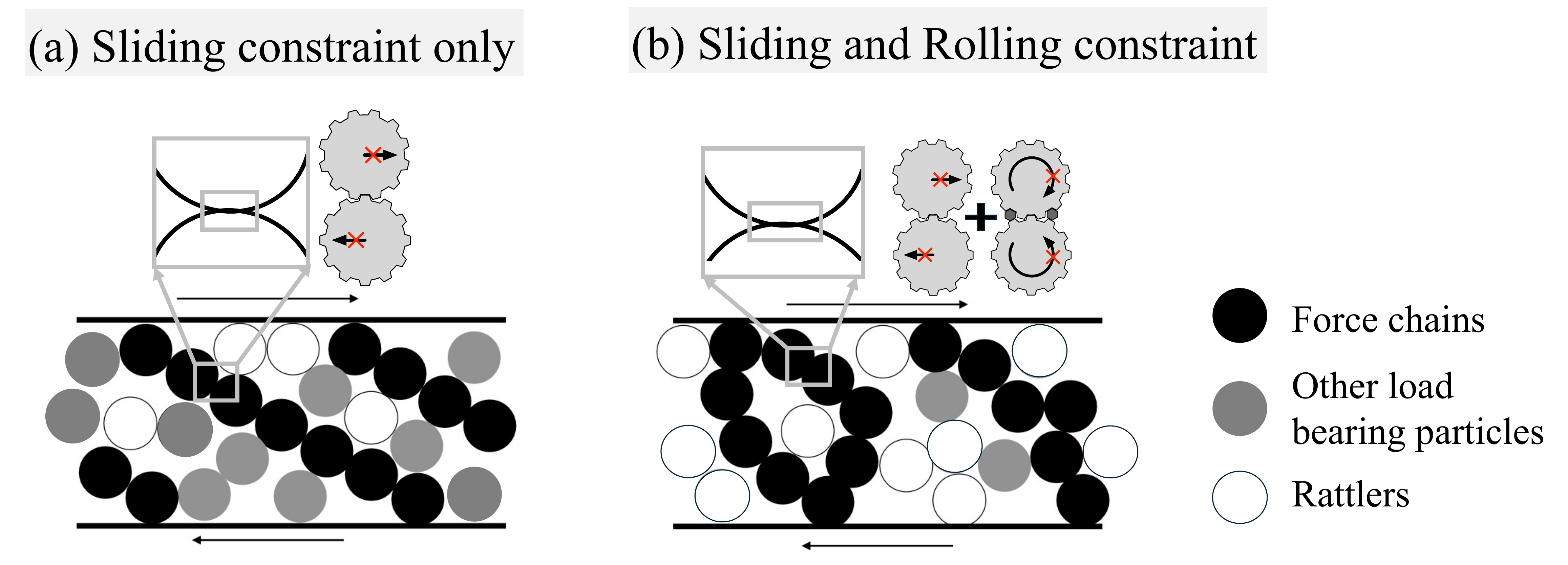}
    \caption{\textbf{Schematic illustration of mechanical stability of packing under shear.} Microstructures and force chain arrangement of particles with (a) sliding constraints only (after Cates \textit{et al.} \cite{Cates_1998a}). The black particles are load-bearing particles forming a force chain; gray particles are other load-bearing particles; white particles are rattlers not participating in a frictional contact network. For particles interacting via sliding friction only, the force chains are collinearly arranged while requiring orthogonal support by other load-bearing particles.
    (b) The proposed mechanical stability diagram for particles interacting with both sliding and rolling constraints. The load-bearing particles do not need orthogonal support, leading to longer, linear force chains.
    The black particles are part of the primary force chain (load-bearing particles) 
    }
    \label{fig:schematic}
\end{figure}

Although mean field models incorporating sliding and rolling constraints predict lower volume fractions for DST, in accordance with several experimental studies~\cite{Hsu_2018, Hsiao_2017, Pradeep_2021}. However, by construction, such approaches are agnostic to the microstructural mechanisms underlying this reduction. \textcolor{black}{ Cates~\textit{et al.}~\cite{Cates_1998a} postulated that a granular packing for mechanical stability under external deformation requires that the SJ state have primary chains (load-bearing contacts) along the compression axis, supported by secondary chains that need to be at least partially in the tensile direction (orthogonal to the compressive axis) within the shear plane.}
We sketch this in Fig.~\ref{fig:schematic}(a); simulations in dense suspensions considering lubrication and frictional forces show a remarkable similarity to this proposition~\cite{Mari_2014, Seto_2013a, thomas2020investigating, Sedes_2022, Sedes_2020, Gameiro_2020, Nabizadeh_2022}. The force chains in the low viscosity state are mainly aligned along the compression axis, with loop-like structures in the DST regime, and eventually the high--viscosity state is stabilized by frictional forces along both the compressive and tensile directions~\cite{Gameiro_2020, Nabizadeh_2022, d2025topological}. Cates \textit{et al.}~\cite{Cates_1998a} noted that orthogonal support is required to provide stability against external perturbation that would cause force chains to buckle. This means that the local relative rotational motion of the particles would cause the force chains to collapse without orthogonal support. This raises an important fundamental question: What is the mechanically stable configuration of particles with both sliding constraints? In other words, does a force chain network formed by rough/faceted particles need orthogonal support as postulated by Cates \textit{et al.} \cite{Cates_1998a}?
We hypothesize that in such systems, force chains may be stable with collinear alignment and without orthogonal support, leading to fewer frictional contacts per particle and a less branched contact network for force transmission, as depicted in Fig.~\ref{fig:schematic}(b).

To test our hypothesis, we apply network science tools to simulation data obtained from a well-established simulation approach for shear-thickening suspensions, considering both sliding and rolling constraints~\cite{Singh_2020, Singh_2022}. Previous studies in dense suspensions~\cite{d2025topological,aminimajd2025scalability,goyal2024flow, Nabizadeh_2022, Sedes_2022, Sedes_2020, Naald_2024, Gameiro_2020} have used network topology tools and successfully correlated the macroscopic bulk response to the mesoscale frictional force and contact networks. Many of these studies have, in turn, been inspired by the usage of such tools in other particulate systems such as dry granular materials~\cite{papadopoulos2018network,bassett2015extraction,tordesillas2007force,tordesillas2010evolution,walker2010topological}. However, to the best our knowledge, none of these studies (especially in dense suspensions) have analyzed the frictional network formed with particles with sliding and rolling constraints. 
Our findings demonstrate that the proposed quantitative characterization method effectively distinguishes distinct frictional contact networks formed by different constraints between particles.
In addition, our network analysis elucidates the underlying factors contributing to the reduction in the volume fractions of DST and SJ as the roughness of the particles increases. Overall, we show that two systems with similar bulk viscosity, i.e., the same distance from their respective jamming points $\phi_J^{\{\mu_s,\mu_r\}}-\phi$ exhibit distinct frictional networks - crucial detail that the mean-field models miss.

\textit{Simulating dense suspensions-} Although real-world systems are three-dimensional, our recent study shows that the main rheological characteristics of a dense suspension under simple shear are very similar in two and three dimensions if the packing fraction is correctly scaled with $\phi_J$~\cite{aminimajd2025scalability}. Given this, for  simplicity of visualization and ease of characterization of the network in lower dimensions, we use 2D simulations. Our simulations involve a monolayer of $N = 2000$ bidisperse (radii $a$ and $1.4a$) rigid non-Brownian spherical particles suspended in a density-matched Newtonian fluid of viscosity $\eta_0$. Due to Stokes flow, the equation of motion involves a force balance between hydrodynamic ($\mathbf {F}_{\rm H}$) and purely repulsive contact ($\mathbf {F}_{\rm C}$) forces, i.e., $ \vec{0}=\vec{F}_C + \vec{F}_H$. Hydrodynamic interactions include single-body Stokes drag and two-body near-field lubrication forces. Contact forces are modeled using virtual linear springs~\cite{Cundall_1979}, and we incorporate sliding and rolling frictions~\cite{Singh_2020}. The friction between particles obeys the Coulomb friction law in both the sliding and rolling modes: $|\vec{F}^C_{slid}| \le \mu_s|\vec{F}^C_{n}|$ and $|\vec{F}^C_{roll}| \le \mu_r|\vec{F}^C_{n}|$. In particular, rolling friction resists motion by a torque (and not force). Thus, the rolling friction force $|\vec{F}^C_{roll}|$ is simply a quasi-force proportional to the relative rolling displacement, which is calculated to compute the rolling torque and does not contribute to the force balance. We employ a critical load model where the normal contact force must exceed a threshold $F_0$ to activate sliding and rolling frictions between particles~\cite{Singh_2020, Singh_2022}. This force gives rise to a typical stress scale: $\tilde{\sigma} = \sigma/\sigma_0, \quad 
\tilde{\dot{\gamma}} = \dot{\gamma}/{\dot{\gamma}_0}, \quad 
\eta_r = \eta/\eta_0$, where $\sigma_0 = F_0/6\pi a^2, \quad 
\dot{\gamma}_0 = F_0/6\pi \eta_0 a^2$ (see \textit{Methods} for details).

\begin{figure*}[tb]
    \centering
        \includegraphics[width=0.8\textwidth]{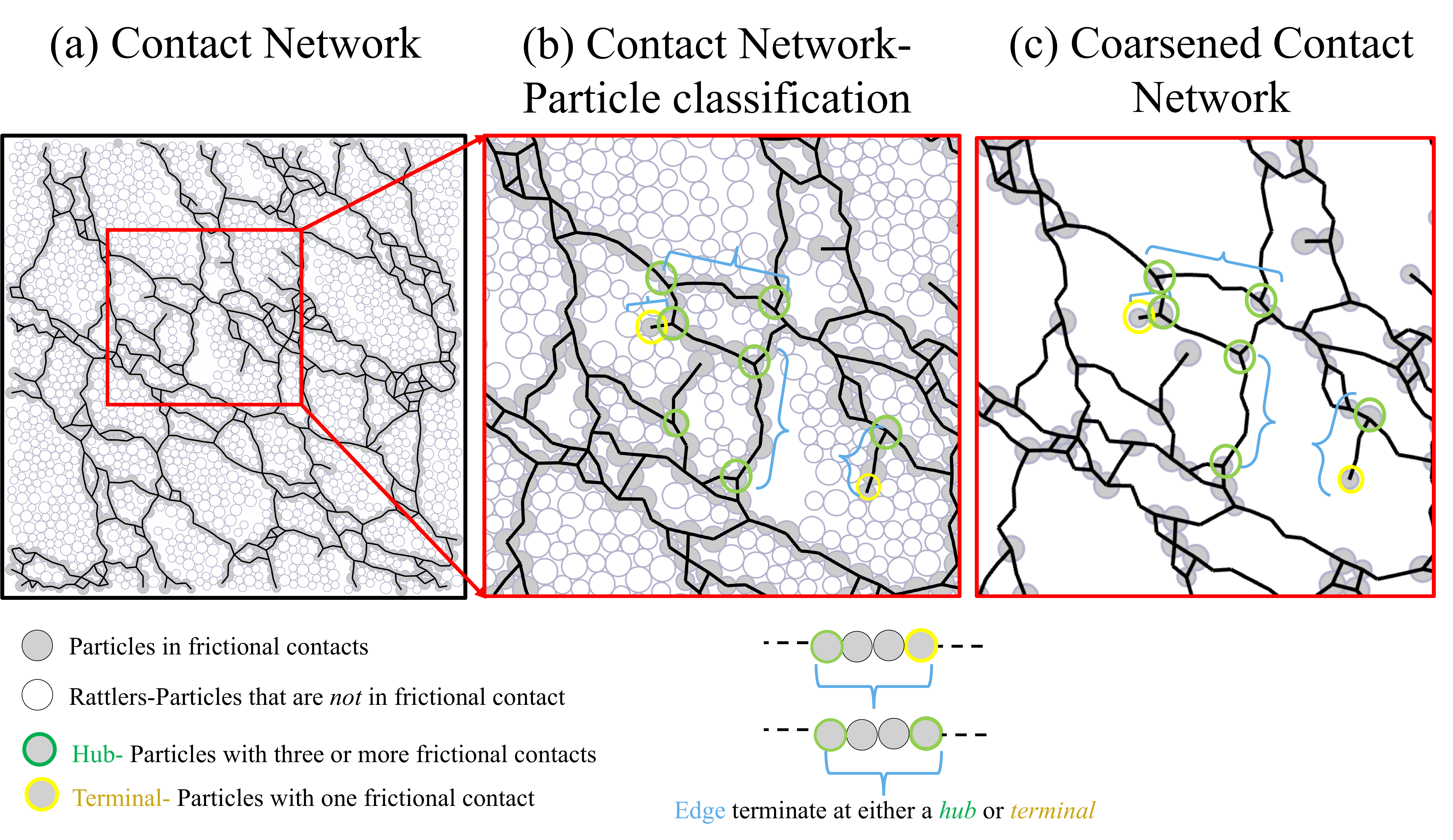}
    \caption{\textbf{Coarsening the frictional contact network and node characterization.} (a) Simulation snapshot showing the frictional contact network, where normal force exceeds the critical threshold, shown in black.(b) A zoomed-in view of the frictional contact network in \textbf{a}, highlighting the hubs, terminals, rattlers, and edges. Rattlers, represented in white, are particles without frictional contact, whereas particles in grey are those in frictional contact. Grey particles are further classified into hubs in green (in frictional contact with three or more particles) and terminals in yellow (in frictional contact with one particle). Edges represent the connections between frictionally contacting particles and terminate at a hub or a terminal. (c) Coarsened contact network illustrating edges connecting hubs, terminals, or a combination of both.}
    \label{fig:definition}
\end{figure*}

\textit{Network Analysis-} To test our hypothesis, we perform a quantitative characterization to identify differences between the two types of contact networks and analyze these differences using a network-based framework. We employ a graph theory approach, the hubs-and-spokes model, to analyze the frictional contact network under shear with different constraints~\cite{papadopoulos2018network,agarwal2010revisiting}. In contrast to previous dry granular studies, we use this model in a slightly different context by analyzing snapshots from simulations in a \textit{dynamic} steady state. The steady-state here is dynamic in the sense that the frictional contacts continuously form, break, and reform as the system is sheared. In the hubs and spokes model, the frictional contact network is coarsened by distinguishing between various degrees of frictional and non-frictional contacts via network analysis. We analyze these coarsened contact networks to identify the nodes responsible for the branching and topology of the network. 

In this approach, we convert the physical system of particles and the contact network into a graph, the particles represented by a node, and the force between particles by an edge, following the methods commonly used in granular systems~\cite{papadopoulos2018network}. Figure \ref{fig:definition}(a) shows a typical frictional contact network obtained from simulations, where the black lines represent the frictional contact network. The particles are then categorized based on their number of contacts. As in dry granular systems~\cite{papadopoulos2018network}, particles with no frictional contacts are identified as rattlers. Particles in frictional contacts with only one particle are labeled terminals, whereas those connected with three or more particles are labeled hubs.  The hubs and terminals define the endpoints of the edges between the particles in frictional contact, collectively forming an edge (or force chains if the edges are weighted, see \textit{Methods} and \textit{Supplementary Information}) as shown in Fig.~\ref{fig:definition}(b) which represents a part of the network shown in Fig.\,\ref{fig:definition}(a). The hubs and terminals act as critical bridging and peripheral nodes in the contact network and participate in the mechanical stability of the packing under external deformation.

To analyze the branching and distribution within the percolating contact network, we coarsen the network into a simplified structure composed of hubs, terminals, and edges connecting these particles. An example of this coarsening process is illustrated in Fig.~\ref{fig:definition}(c), which shows the coarsened form of the contact network shown in Fig.~\ref{fig:definition}(b). Finally, we analyze network connectivity by measuring the edge lengths corresponding to the force chain lengths ($\mathcal{L}$) and evaluate their alignment with the compressive direction $(\theta)$. The angles between consecutive particles that form an edge are used to determine the linearity of the force chains, as shown in Fig.~\ref{fig:definition}(c).

\begin{figure*}[tb]
    \centering
        \includegraphics[width=1.05\textwidth]{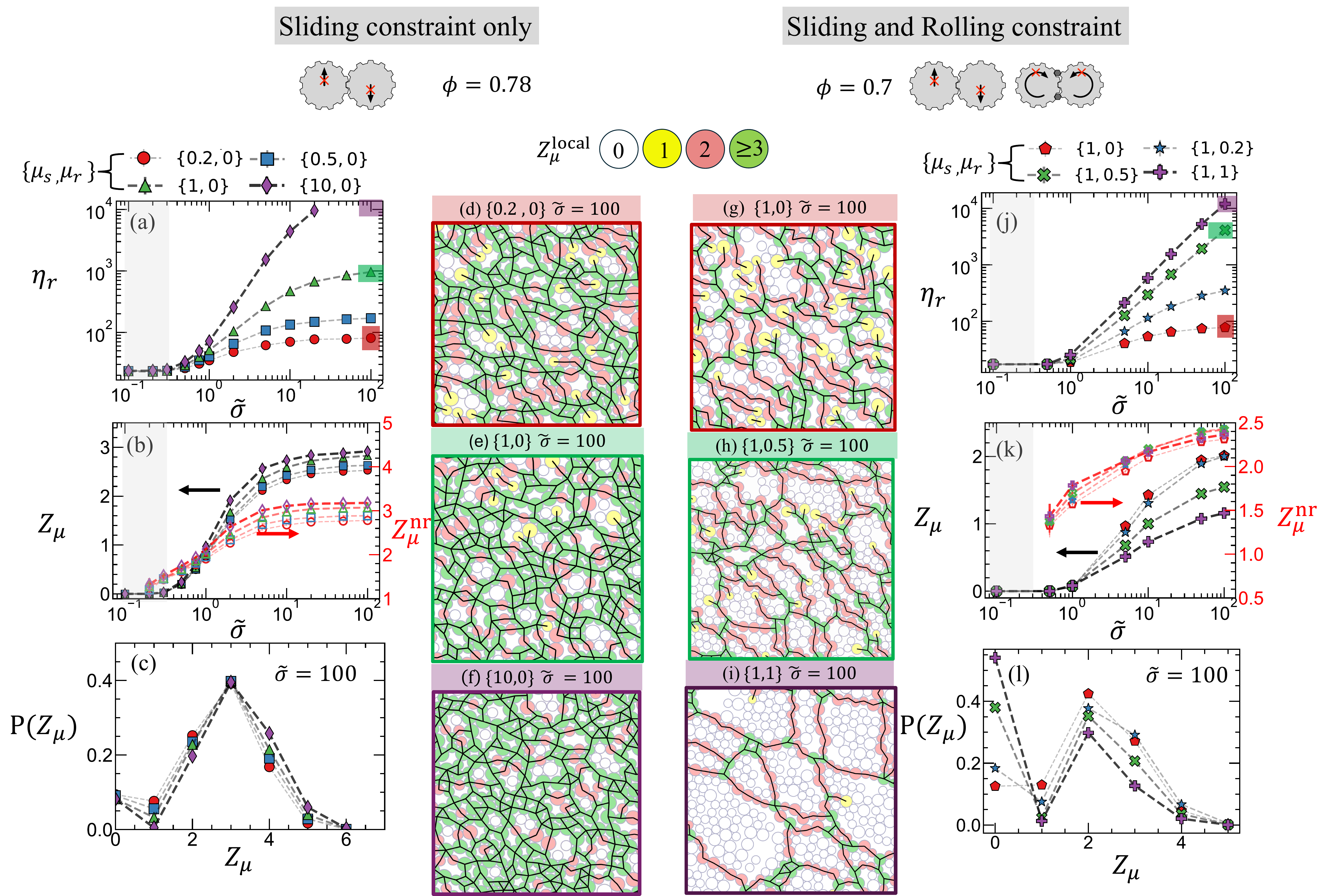}
    \caption{\textbf{Rheology, frictional contact network and coordination number.}
    Flow curves (a, j), frictional coordination numbers (b, k) for total $(Z_{\mu})$ and non-rattlers $(Z_{\mu}^{\mathrm{nr}})$, probability distribution functions of $(Z_{\mu})$ at $\tilde{\sigma}=100$ (c, l), and microstructural snapshots of dense suspensions undergoing continuous and discontinuous shear thickening, as well as approaching shear jamming (d–i). The snapshots depict suspensions composed of particles exhibiting only sliding friction (d–f) and particles with both sliding and rolling frictions (g–i). Figures (a–f) present the microstructural analysis of a dense suspension constrained by sliding friction at a volume fraction of 0.78, under varying shear stress ($\tilde{\sigma}$) and sliding friction coefficient ($\mu_s$), with rolling friction absent ($\mu_r = 0$). Zoomed-in microstructure snapshots (d–f) illustrate the thickened states for continuous shear thickening (CST, $\mu_s = 0.2$, $\tilde{\sigma} = 100$), discontinuous shear thickening (DST, $\mu_s = 1$, $\tilde{\sigma} = 100$), and approaching shear jamming ($\mu_s = 10$, $\tilde{\sigma} = 100$). Figures (g–l) display the microstructural analysis of a dense suspension at a volume fraction of 0.7, constrained by sliding friction under varying shear stress ($\tilde{\sigma}$) and rolling friction coefficient ($\mu_r$), with a fixed sliding friction coefficient ($\mu_s = 1$). The zoomed-in snapshots (g–i) illustrate the thickened states for CST ($\mu_r = 0$, $\tilde{\sigma} = 100$), DST ($\mu_r = 0.5$, $\tilde{\sigma} = 100$), and approaching shear jamming ($\mu_r = 1$, $\tilde{\sigma} = 100$). Particles are color-coded according to their frictional coordination number ($Z_\mu$), with white particles representing rattlers. Black lines indicate frictional contacts forming force chains. \textcolor{black}{The shaded region in (a), (b), (j), (k) shows the range of shear stress for which frictional interactions are not in effect in the simulations.} }
    \label{fig:snapshots}
\end{figure*}

 \textit{Rheology, coordination number, and microstructure-} 
We now present a sweep of constraints $\{\mu_s,\mu_r\}$ at two distinct packing fractions $\phi=0.7$ and 0.78; these roughly correspond to 3D volume fractions of 0.5 and 0.56, respectively. For $\phi=0.78$, we vary $\mu_s = [0.2, 0.5, 1, 10]$ while keeping $\mu_r=0$; while in the case of simulations at $\phi=0.7$, $\mu_s=1$ is maintained changing the rolling friction $\mu_r = [0, 0.2, 0.5, 1]$. 
Figure~\ref{fig:snapshots} presents key rheological and microstructural characteristics, including flow curves, network visuals, frictional coordination number (total and non-rattlers) and probability distribution function $ P(Z_\mu ( \tilde{\sigma} = 100)$.

Figures \ref{fig:snapshots}a and \ref{fig:snapshots}j show the typical shear-thickening behavior observed in dense suspensions, that is, viscosity increase between a low- and high-viscosity plateau. With an increase in constraints $\{\mu_s,\mu_r\}$, the shear-thickening behavior transitions from CST to DST and eventually to SJ. For low friction, the shear-thickening is continuous (CST). As ${\mu_s, \mu_r}$  increases, shear-thickening becomes discontinuous, as evidenced by $\eta_r \propto \sigma/\sigma_0$. Figures S1-S2 of the Supplementary Information show $\eta_r(\dot{\gamma})$, where the viscosity becomes non-monotonic as the constraints are increased, denoting DST. In our simulations, since the friction is stress-activated, at low stresses $\tilde{\sigma} \ll 1$, the suspension is in the lubricated frictionless state and $\eta_r$ is independent of the friction coefficient. This is also indicated by $Z_\mu=0$ (Figs.~\ref{fig:snapshots}b and k). Increasing the applied stress $\tilde{\sigma}$ at fixed $\phi$ leads to direct frictional contacts, and the system transitions to the frictional state at $\tilde{\sigma} \gg 1$. The nature of the viscosity increase (CS/DST/SJ) is then simply given by the position of $\phi$ relative to the frictional jamming point $\phi_J^{\{\mu_s,\mu_r \}}$: the more constraints are added, the stronger the shear thickening. The results for the first normal stress difference $N_1/\tilde{\sigma}$ are shown for both cases in Fig.\, S3 of Supplementary Information, where $N_1/\tilde{\sigma}$ is found to be larger for large $\mu_r$.

To provide a visual insight - at the particle level - on how constraints affect the frictional contact network, we show (Figs.~\ref{fig:snapshots}d–f and \ref{fig:snapshots}g-i) representative snapshots of the frictional contact network (\textcolor{black}{full snapshots of the frictional contact network is shown in Figs.\,S5-S6 in Supplementary Information for both cases, sliding-only and sliding-and-rolling constraints}). In the sliding-only case $\{\mu_s,0\}$, increasing $ \mu_s $ enhances connectivity, forming a highly interconnected contact network at high $\mu_s$, leading to enhanced viscosity during DST and SJ. In contrast, increasing $\mu_r$ reduces connectivity in the sliding-and-rolling case, even though the system still undergoes DST and SJ. This observation appears \textit{counterintuitive} given that both systems exhibit a comparable macroscopic viscosity but undergo different stress transmission mechanisms, highlighting the ``roll'' of friction. \textcolor{black}{(The effect of interparticle friction on particle orientation relative to the shear direction, highlighting anisotropy in the system, is analyzed through $P(\theta)$ versus $\theta$ for both cases in Fig.,S8 of the Supplementary Information.) } 

Next, to provide a quantitative measure, we report the frictional coordination number, both total $Z_{\mu}$ \textcolor{black}{(average number of frictional contacts per particle)} and non-rattler $ Z_{\mu}^{\mathrm{nr}} $ \textcolor{black}{(average number of frictional contacts per non-rattler particle)} (Figures \ref{fig:snapshots}b and \ref{fig:snapshots}k). In the sliding-only case, increasing $\mu_s$ resists the relative sliding motion, thereby increasing both $Z_\mu$ and $Z_\mu^{nr}$~\cite{aminimajd2025robust, aminimajd2025scalability, d2025topological}. This is also consistent with previous 3D observations~\cite{radhakrishnan2019force}. However, the sliding-and-rolling case is more intriguing; $Z_\mu$ and $Z_\mu^{nr}$ increase with $\tilde{\sigma}$, given that friction is stress-activated. However, $Z_\mu$ decreases as $\mu_r$ increases for a constant $\mu_s$. This is consistent with visual observations of the frictional contact network that, with a sufficiently large $\mu_r$, only a few particles are needed for mechanical stability, leading to a large number of rattlers (Fig.~\ref{fig:snapshots}g).

Finally, the contact number distribution ($ P(Z_{\mu}(\tilde{\sigma}=100)) $) further reinforces these findings. In the sliding-only case, increasing $ \mu_s $ shifts the probability distribution towards higher $ Z_{\mu} $, strengthening the contact network while maintaining the peak at $Z_\mu=3$. In the sliding-and-rolling case, increasing $\mu_r$ drastically affects $P(Z_\mu)$. $P(Z_\mu=0)$ increases with $\mu_r$, which reflects the increase in the number of rattlers. Although the peak position in $P(Z_\mu)$ is insensitive to $\mu_r$, the peak value decreases with $\mu_r$. In particular, the peaks in the sliding and rolling case changes to $ Z_{\mu} = 2 $ for the sliding-and-rolling friction system, consistent with the saturated $ Z_{\mu} $ observed in Figs.~\ref{fig:snapshots}b and \ref{fig:snapshots}k.

These findings highlight a fundamental distinction in how the two systems approach DST and SJ. In the sliding-only case, DST and SJ occur through the formation of a highly interconnected force network, where a larger number of particles actively participate in stress transmission. In contrast, in the sliding-and-rolling case, DST and SJ occur through a less interconnected and \textit{highly} heterogeneous contact network, requiring fewer stress-bearing particles (and more rattlers are present).
 Despite their similar macroscopic viscosity, their microstructural evolution and force transmission mechanisms differ significantly, which we explore next.

\begin{figure*}[t]
    \centering
        \includegraphics[width=1.05\textwidth]{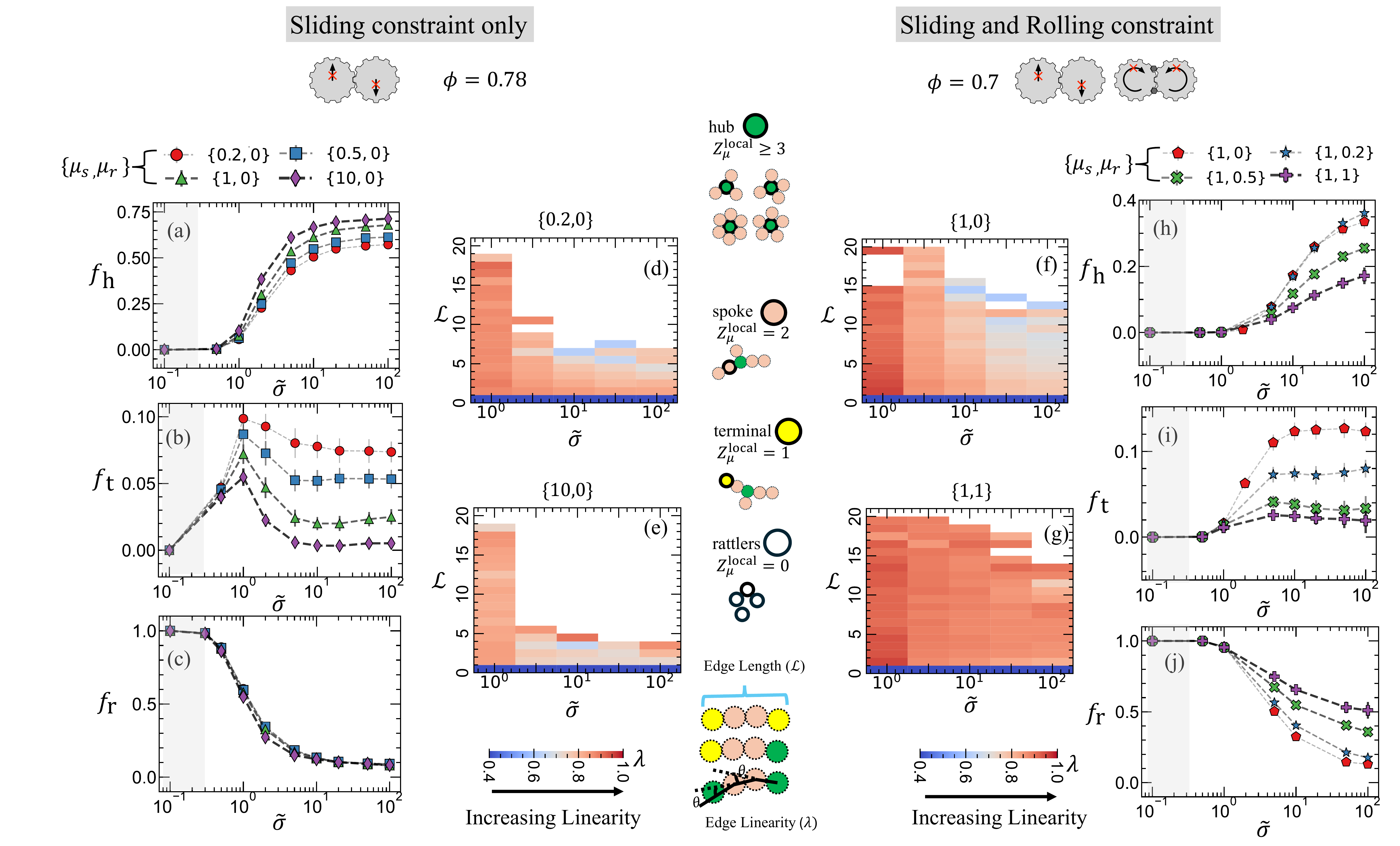}
    \caption{\textbf{Hubs-and-spokes model based insights into frictional contact network.} Statistical metrics as functions of shear stress ($\tilde{\sigma}$): fraction of hubs $(f_{\mathrm{h}}(Z_{\mu}^{\mathrm{local}}\geq3))$ (a, h), fraction of terminals $(f_{\mathrm{t}}(Z_{\mu}^{\mathrm{local}}=1))$ (b, i), fraction of rattlers $f_{\mathrm{r}}(Z_{\mu}^{\mathrm{local}}=0)$ (c, j), and edge length $\mathcal{L}$ and linearity of force chains $\lambda$ (d–g). \textcolor{black}{$(Z_{\mu}^{\mathrm{local}}$ is the local count of frictional contact per particle).} \textcolor{black}{(The variation of fraction of spokes ($f_S$, where $f_h+f_s+f_t+f_r=1$) as a function of interparticle friction in both cases, sliding -only and sliding-and-rolling, is shown in Fig.\,S4 of Supplementary Information.)} The microstructural analysis is performed on dense suspensions undergoing CST, DST, and approaching shear jamming. Two different cases are considered: Figures (a–e): A dense suspension constrained by sliding friction at a volume fraction of 0.78, with varying shear stress ($\tilde{\sigma}$) and sliding friction coefficient ($\mu_s$), while rolling friction is absent ($\mu_r = 0$). Figures (f–i): A dense suspension constrained by both sliding and rolling friction at a volume fraction of 0.7, with varying shear stress ($\tilde{\sigma}$) and rolling friction coefficient ($\mu_r$), while the sliding friction coefficient is fixed at $\mu_s = 1$. \textcolor{black}{The shaded region in (a), (b), (c), (h), (i), (j) shows the range of shear stress for which frictional interactions are not in effect in the simulations.} }
    \label{fig:edge_rattlers}
\end{figure*}

\textit{Contact network analysis} - To quantitatively track the differences in the frictional contact network formed with different combinations of $\{\mu_s,\mu_r\}$, we employ the hubs-and-spokes model, as introduced in Fig.~\ref{fig:definition}. Hubs and terminals act as critical bridging and peripheral nodes in the contact network and participate in the mechanical stability of the packing under external deformation. \textcolor{black}{The local count of the number of frictional contacts can also be represented as $Z_{\mu}^{\mathrm{local}}$. Therefore, $Z_{\mu}^{\mathrm{local}}\geq3$ for hubs, $Z_{\mu}^{\mathrm{local}}=2$ for spokes, $Z_{\mu}^{\mathrm{local}}=1$ for terminals, and $Z_{\mu}^{\mathrm{local}}=0$ for rattlers. We also measure the length of the edge $\mathcal{L}$ and the linearity of the edge $\lambda$. Here, the edge length $\mathcal{L}$ is defined as the number of intermediate connections between the particles in an edge, including the hubs or terminals at each end, and the intermediate spokes and edge linearity $\lambda$ is the average cosine of the angles between consecutive particles in an edge, indicating the extent of straightness or closeness to the complete linearity of the edge. }

In the sliding-only case, the contact network is highly branched, and the fraction of hubs $f_h$ (Fig.~\ref{fig:edge_rattlers}(a)) increases monotonically with both $\tilde{\sigma}$ and $\mu_s$ reflecting the observation. However, the fraction of terminals $f_t$ varies non-monotonically with $\tilde{\sigma}$; it increases initially with stress for $\tilde{\sigma} \le 1$ and decreases upon further increase in stress. The initial increase is insensitive to $\mu_s$, while the fraction of terminals $f_T$ decreases with $\mu_s$ for intermediate to high stress values ($\tilde{\sigma} > 1$). This decrease corresponds to the emergence of closed structures (i.e., loops) at these stress levels, as recently reported~\cite{d2025topological}. In contrast, in the sliding-and-rolling case, the fraction of hubs $f_\mathrm{h}$ decreases with increasing $\mu_r$ (Fig.~\ref{fig:edge_rattlers}h). The fraction of terminals $f_\mathrm{t}$ increases with $\tilde{\sigma}$ and saturates for $\tilde{\sigma} \ge 5$ and does not show a pronounced peak, as in the sliding-only case. In particular, the fraction of terminals $f_\mathrm{t}$ remains approximately of the same order in both cases, while the fraction of hubs $f_\mathrm{h}$ is approximately two times higher in the sliding case than the sliding-and-rolling case.

Next, we examine the fraction of rattlers (particles with zero frictional contact) as shown in Fig.~\ref{fig:edge_rattlers}(c) for the sliding-only case and Fig.~\ref{fig:edge_rattlers}(j) for the sliding-and-rolling case. The fraction of rattlers decreases with $\tilde{\sigma}$ in both cases, as expected. This is roughly insensitive to an increase in $\mu_s$ (Fig.~\ref{fig:edge_rattlers}c). However, the fraction of rattlers increases with increasing $\mu_r$. This result is striking, as by increasing the constraints $\{\mu_s,\mu_r\}$ increases the viscosity of the suspension, and the suspension undergoes DST and SJ.  In this case, fewer particles participate in the frictional contact network, yet the shear-thickening behavior is enhanced. \textcolor{black}{(The variation in the fraction of spokes as a function of interparticle friction in both cases, sliding -only and sliding-and-rolling, is shown in Fig.\,S4 of Supplementary Information.)}

Thus far, the analysis indicates that the number of hubs and particles that do not participate in the friction network (rattlers) increases with $\mu_r$. We assume that this affects the length of the force chains. To verify this, we next analyze the edge length\textcolor{black}{, ($\mathcal{L}=
 1 + \sum_{i=1}^{N-1}n_{\text{spokes,i}}$)} and edge linearity\textcolor{black}{, ($\lambda=
\frac{1}{N-2} \sum_{i=1}^{N-2} \cos\theta_i$)} functions of stress $\tilde{\sigma}$ for various combinations of interparticle friction. This is presented in Figs.~\ref{fig:edge_rattlers}(d-for $\mu_s=0.2$)-(e-for $\mu_s=10$) for the sliding-only case and Figs.~\ref{fig:edge_rattlers}(f-for $\mu_r=0$)-(g-for $\mu_r=1$)for the sliding-and-rolling case.  Both edge length and linearity of the edge are analyzed using the probability distribution function of edge length for different $\tilde{\sigma}$ and the linearity of each edge is highlighted using the color scheme of each bar utilizing the colorbar (cooler less linear and warmer closer to linear), as shown in Fig.\,\ref{fig:edge_rattlers}(d). \textcolor{black}{For the sliding-only case, edge lengths with a nonzero probability distribution function are represented by colored cells. We observe that as the stress increases, the range of these colored cells narrows, indicating smaller edge lengths in both cases. However, in the sliding-and-rolling case, the colored cell range is broader, capturing both smaller and larger edge lengths. Additionally, for the sliding-only case, the colors transition to cooler shades as the sliding constraint increases from $\mu_s=0.2$ to 10, as shown in Fig.~\ref{fig:edge_rattlers}(d-e). In contrast, for the sliding-and-rolling case, the colored cells appear relatively warmer for $\mu_r=1$ (Fig.~\ref{fig:edge_rattlers}g) than $\mu_r=0$ (Fig.~\ref{fig:edge_rattlers}f).
 }

In the sliding-only case, the frictional network is more branched and is composed of frictional contacts in both principal directions, hinting at smaller edges, which we observe in Fig.~\ref{fig:edge_rattlers}d. However, we do not observe a clear trend in our analysis for edge linearity (Fig.~\ref{fig:edge_rattlers}e). In contrast, for the sliding-and-rolling case, the rolling constraint can stabilize the long force chains, as depicted in Fig.~\ref{fig:snapshots}. Figures~\ref{fig:edge_rattlers}f-g confirm our hypothesis, the edge length increases significantly with $\mu_r$. Additionally, while shorter edges do not show a specific trend toward linearity, rolling friction causes longer edges to become more linear and remain stable, particularly for large $\mu_r$. \textcolor{black}{(The effect of interparticle friction on the probability distribution function of edge length for $\tilde{\sigma}=100$ is shown in Figs.\,S7 in the Supplementary Information for both cases.)}

The results presented thus far lead us to propose a different picture of the mechanical stability of the force chain arrangement for particles with sliding and rolling constraints, as shown in Fig.~\ref{fig:schematic}b. This new proposition has key differences from that of Cates \textit{et al.}~\cite{Cates_1998a}. Cates \textit{et al.} proposed that the load-bearing particles participating in the force chain network need orthogonal support from the other load-bearing particles; in the absence of these particles, the force chains will buckle under external deformation. Our simulations in the sliding-only case corroborate the scenario proposed by Cates \textit{et al.}~\cite{Cates_1998a}. Before discussing our proposition, we explore Cates's proposition within our framework. Upon coarsening the force chain arrangement as proposed by Cates \textit{et al.}, it exhibits a large fraction of hubs (particles in frictional contact with three or more particles) with a lower fraction of terminals (particles in frictional contact with only one particle) and a very few (yet finite) fraction of rattlers. This also implies that the force chains are shorter because of branching for such nominally rough particles (sliding-only case). In contrast, rough or faceted particles with sliding-and-rolling constraints can withstand external deformation even without orthogonal support. In the coarsened network, this appears as a large fraction of rattlers along with fewer bridging particles (hubs) and terminals, thus resulting in linear and longer force chains. 

Finally, although rheology is emerging as a lens for microscopic physics~\cite{Singh_2022}, mean field models suggest that the distance from jamming should lead to the same rheology in terms of viscosity or strength of shear-thickening~\cite{Pradeep_2021, Singh_2020, Wyart_2014}. However, we show that the stress propagation mechanism or mesoscale physics for suspensions, even with the same viscosity, can be \textit{strikingly} different and depends sensitively on the particle-level details such as sliding and rolling frictions. Our analysis also provides a mechanistic explanation for the reduction in the DST and SJ volume fraction observed for rough particles. Due to their tendency to form a longer-linear force chain arrangement due to resistance to torque or rolling over, rough particles exhibit DST and shear jamming with significantly fewer particles actively participating in the force chain or contact network formation at reduced packing fractions. This highlights the distinct role of particle roughness in altering the structural and mechanical characteristics of the DST and shear-jammed states.

\textit{Conclusions}- In this study, we focus on analyzing the frictional contact network since this sub-network is the driver of strong shear thickening and shear jamming~\cite{Seto_2013a, Mari_2014, Singh_2020, Gameiro_2020, Naald_2024, Nabizadeh_2022, Clavaud_2017, Hsu_2018}. We particularly focused on the stress transmission mechanism across nominal frictional and highly rough particles, i.e., particles with only sliding and both sliding and rolling frictions, respectively. 
Using our hub-and-spoke model, we quantified the frictional contact network and correlated the mesoscale network features with the bulk response of the suspension. 

It has been well established in the literature that the roughness of the particle level can reduce discontinuous shear thickening (DST) that occurs at relatively lower volume fractions~\cite{Lootens_2005, Hsiao_2017, Hsu_2018}. Recent studies suggested that interlocking between the particle surfaces due to large asperities would hinder relative rolling between particles together with a sliding motion that leads to a lower jamming volume fraction $\phi_J^{\{\mu_s,\mu_r\}}$. This, in spirit, of the mean-field models suggests a reduced distance from jamming ($\phi_J^{\{\mu_s,\mu_r\}}-\phi$), thus reducing the volume fraction of the onset for DST $\phi_{\mathrm{DST}}$. This assumes that stress transmission is unaffected. However, we show that rough particles exhibit a lower DST and jamming volume fraction because they tend to form linear, longer, and less branched force chains.

Our findings are consistent with previous results for dense amorphous materials, especially granular materials, dense suspensions, and colloidal gels. In granular materials, load-bearing particles (or strong force chains~\cite{Radjai_1998}) are known to be the main contributors to stress and align along the compressive axis, with orthogonal support provided by other load-bearing particles and rattlers~\cite{Cates_1998a, Radjai_1998,Silke_2016}. Dense suspensions with nominally frictional particles exhibit a very similar force network structure that leads to the emergence of closed ``loop''-like structures that are critical for DST~\cite{Gameiro_2020, d2025topological}, which is also supported by recent experimental evidence~\cite{Pradeep_2021}. We await experimental guidance in confirming our picture of mechanical stability for rough or faceted particles that restrict both relative sliding and rolling motion between particles~\cite{scherrer2024sliding, Hsu_2018,d2023role, Pradeep_2021}.  Although our simulations are conducted in 2D, the techniques and methods used here are, in principle, transferable to 3D systems, \textcolor{black}{but can be computationally more expensive for 3D systems}. However, the transition to 3D introduces additional challenges, such as accounting for both planar and out-of-plane \textcolor{black}{force chains, linearity of force chains}, which require further consideration. We envision that this approach can provide insight into a wide range of flowing amorphous systems such as emulsions and gels, especially those composed of particles where constraints on the relative motion of the particles have recently been used to manipulate the bulk response~\cite{muller2023toughening,müller2025tuning,workamp2019contact,richards2019role}.

\newpage
\clearpage

\section*{Methods} We simulate the simple-shear flow of a monolayer of $N = 2000$ bidisperse non-Brownian frictional rigid spheres immersed in a Newtonian fluid under imposed stress using Lees-Edwards boundary conditions~\cite{Lees_1972, Mari_2015}. To avoid particle ordering, spheres with radii $ a $ and $ 1.4a $ are mixed in equal volume. Particles interact via lubrication forces and Coulomb frictional contacts, reproducing the experimentally observed shear thickening~\cite{Singh_2020, Singh_2022, Mari_2015}.

Assuming inertialess motion ($ \text{Reynolds number} = 0 $), the system satisfies the force balance between the lubrication forces ($ \vec{F}_H $) and the contact ($ \vec{F}_C $) forces: $ \vec{0} = \vec{F}_H(r, u) + \vec{F}_C(r)$, where $ u $ and $ r $ are the velocity and position vectors ($ u \equiv \dot{r} $). The lubrication resistance is truncated at $ h = 10^{-3}a $, and contacts are allowed. The contact forces follow the Coulomb friction law $|\vec{F}^C_{slid}| \le \mu_s|\vec{F}^C_{n}|$ and $|\vec{F}^C_{roll}| \le \mu_r|\vec{F}^C_{n}|$. $ F_C $ represents the contact force, which includes the repulsive normal contact force ($ F^C_N$), the tangential Coulombic friction force ($ F^{C}_{slid} $) and the rolling friction contact force $ F_{roll}^{C} $. Here, $ F^{C}_{roll} $ is a quasi-force used to calculate the torque. A critical load model (CLM) introduces a threshold force $ F_0 $ required to activate friction. The implementation of $F_{roll}^{C}$ is discussed in detail in previous studies~\cite{Singh_2020, Singh_2022}.

Since the simulations are performed under an imposed fixed shear stress $\sigma$, suspension flows with a fluctuating time-dependent shear rate $\dot{\gamma}$. The data presented in this study are averaged in the steady state after omitting transients that last approximately $\mathcal{O}(1)$ strain units. The relative viscosity is calculated as $\eta_r(t) = \sigma/\eta_0\dot{\gamma}(t)$, with $\eta_0$ being the liquid viscosity.
The critical load force $ F_0 $ establishes a stress scale $ \sigma_0 = F_0/6\pi a^2 $, roughly indicating the onset of the shear thickening. Throughout this study, the quantities are presented in terms of the dimensionless stress $ \tilde{\sigma} =\sigma/\sigma_0 $ and scaled strain rate $\dot{\gamma}/\dot{\gamma}_0 $, where $ \dot{\gamma}_0 = F_0/6\pi \eta_0 a^2 $. 

The simulations calculate the positions of the particles, normal and tangential contact forces, and non-contact lubrication forces. The physical space is converted to a graph, with the positions of the particles and frictional contacts represented as nodes and edges, respectively. This allows the construction of an adjacency matrix to characterize the evolving contact network. Particles are categorized as hubs (frictional contact with three or more neighbors), terminals (contact with one neighbor), \textcolor{black}{spokes (frictional contact with two neighbors)}, or rattlers (no frictional contacts). The adjacency matrix includes edge weights based on the magnitude of frictional or normal contact forces, as shown in Figs.~S5–S6 in the Supplementary Information. For all other results presented in Figs.~\ref{fig:snapshots}-\ref{fig:edge_rattlers}, the edges are treated as unweighted to focus on characterizing the contact network. We calculate the frictional coordination number (\textcolor{black}{ average number of frictional contacts per particle}, $Z_\mu$\textcolor{black}{=2(total number of frictional contacts)/(total number of particles)}), its non-rattler counterpart ( \textcolor{black}{(average number of frictional contacts per non-rattler particle)}, $Z_\mu^{\mathrm{nr}}$\textcolor{black}{=2(total number of frictional contacts)/(total number of particles-total number of rattlers)}), and the probability distribution function of $Z_\mu$, \textcolor{blue}{P($Z_\mu$)}. In addition, the edge linearity, edge lengths, and PDF of normal contact forces are analyzed to quantify the force chain topology. The temporal evolution of the contact network is analyzed to track the formation, evolution, and reorganization of the force chains under imposed shear.


The force chains are identified by coarsening the force chain network by removing all nodes with exactly two connections and connecting their adjoining neighbors. The motivation is that these nodes pass stress along a chain and thus comprise the force chains that we wish to identify. Performing this action sequentially on all nodes results in a coarsened network that largely comprises high-degree nodes, which we call hubs \textcolor{black}{$Z_{\mu}^{\mathrm{local}}\geq3$}, which are the key points of stress propagation. The nodes with exactly one contact are called terminals \textcolor{black}{$Z_{\mu}^{\mathrm{local}}=1$}, \textcolor{black}{$Z_{\mu}^{\mathrm{local}}$ is the count for the local frictional contact of a particle}~(\textcolor{black}{the nodes with two frictional contact are called spokes \textcolor{black}{$Z_{\mu}^{\mathrm{local}}=2$} and the nodes with no frictional contact are called rattlers \textcolor{black}{$Z_{\mu}^{\mathrm{local}}=0$}}). Connecting these hubs and terminals are new edges that represent individual force chains. We keep track of the nodes that were removed for each force chain for further analysis. Force chain members also include connected hubs and terminals. This coarsening allows us to characterize the properties of force chains directly, such as the force chain length, \textcolor{black}{($\mathcal{L}=
 1 + \sum_{i=1}^{N-1}n_{\text{spokes,i}}$, $n_\text{spokes,i}$ is the count for each spoke, $N$ is the total number of spokes)}, defined as the number of intermediate connections between particles forming an edge 
 and linearity (\textcolor{black}{$\lambda=
\frac{1}{N-2} \sum_{i=1}^{N-2} \cos\theta_i$, $N$ is total number of spokes in an edge}) is characterized by the average deviation angle between subsequent nodes of the chain as shown in \ref{fig:edge_rattlers}). These two parameters can be measured simultaneously to analyze the linearity as a function of the force chain length as the constraints vary.

\bibliographystyle{unsrt} 
\bibliography{dst} 
\newpage
\section*{Supplementary Information}\label{section_SI}
\renewcommand\thefigure{S\arabic{figure}} 
 \setcounter{figure}{0}

In this section we provide details about effect of sliding-only constraints and sliding-and-rolling constraints on (i) rheology of dense suspensions through flow curve, first normal stress difference; (ii) frictional contact network snapshots; (iii) network analysis through fraction of spokes and edge length probability distribution function; and (iv) anisotropy of frictional contact network.

\section{Rheology of dense suspensions}

\subsection{Sliding-only constraints}
\begin{figure*}[h]
    \centering
    \includegraphics[width=0.9\linewidth]{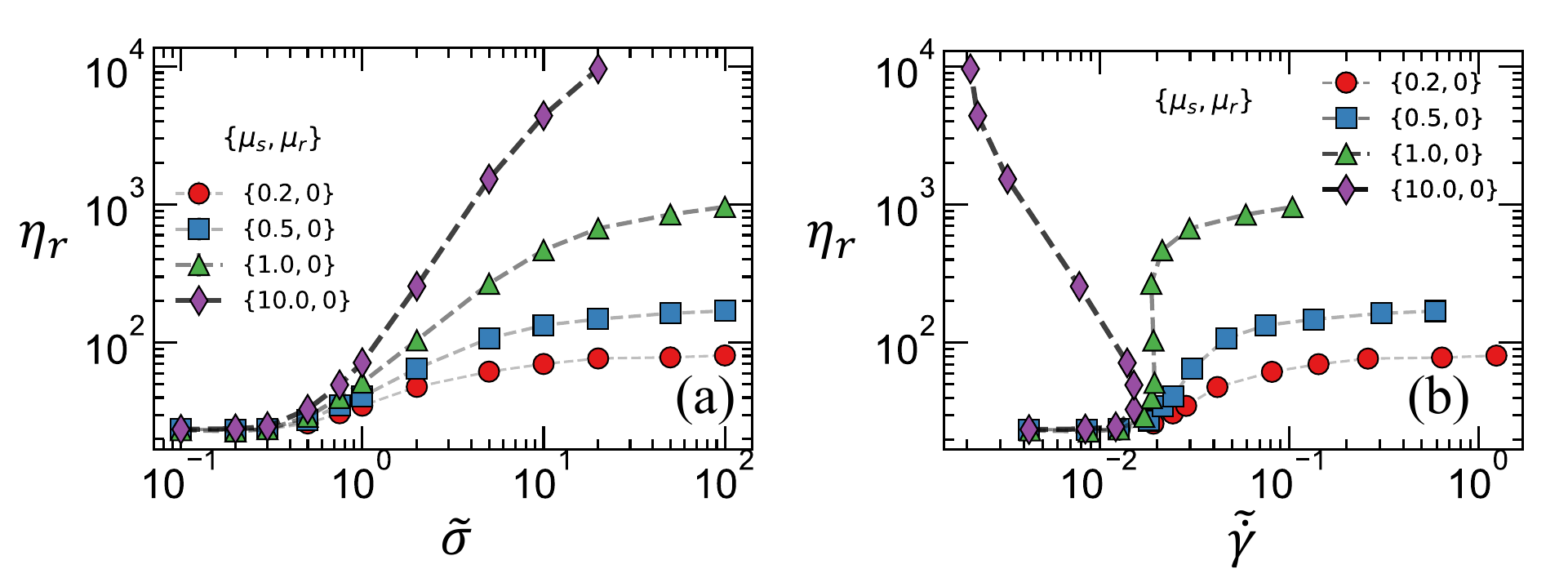}
    \caption{Viscosity $\eta_r$ versus shear stress $\tilde{\sigma}$ (a) and viscosity $\eta_r$ versus shear rate $\tilde{\dot{\gamma}}$ (b) is plotted for dense suspension with sliding constraint only at $\phi=0.78$ for different sliding constraints, $\mu_s$ ranging from 0.2 to 10. The flow curve shows thickened states for CST ($\mu_s = 0.2$, $\tilde{\sigma} = 100$), DST ($\mu_s = 1$, $\tilde{\sigma} = 100$), and approaching shear jamming ($\mu_s = 10$, $\tilde{\sigma} = 100$). }
    \label{fig:flow_curve_combined_sliding}
\end{figure*}
\clearpage
\subsection{Sliding-and-Rolling constraints}
\begin{figure*}[h]
    \centering
    \includegraphics[width=0.9\linewidth]{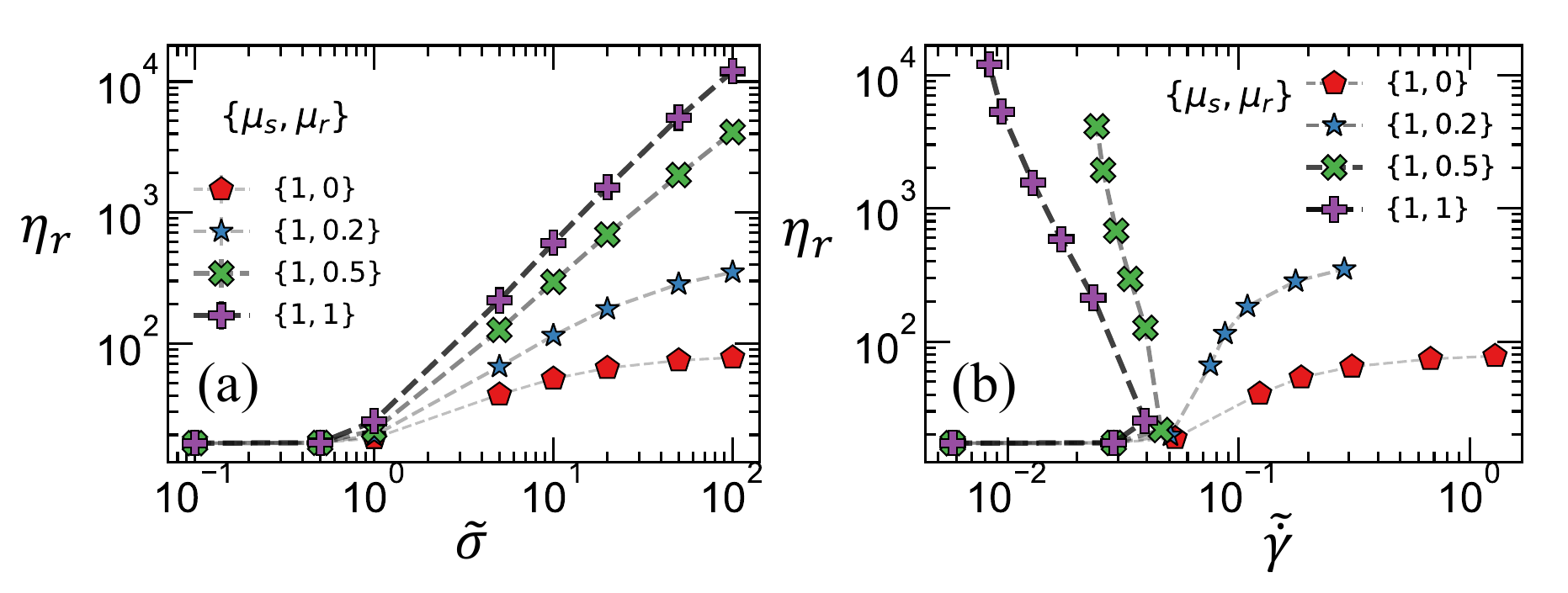}
    \caption{Viscosity $\eta_r$ versus shear stress $\tilde{\sigma}$ (a) and viscosity $\eta_r$ versus shear rate $\tilde{\dot{\gamma}}$ (b) is plotted for dense suspension with sliding constraint only at $\phi=0.7$ for different sliding constraints, $\mu_r$ ranging from 0 to 1.  The flow curve shows thickened states for CST ($\mu_r = 0$, $\tilde{\sigma} = 100$), DST ($\mu_r = 0.5$, $\tilde{\sigma} = 100$), and approaching shear jamming ($\mu_r = 1$, $\tilde{\sigma} = 100$).}
\label{fig:flow_curve_combined_rolling}
\end{figure*}
\clearpage
\subsection{ First normal stress difference in dense suspensions with sliding-only and sliding-and-rolling constraints}

\begin{figure*}[h]
    \centering
    \includegraphics[width=0.9\linewidth]{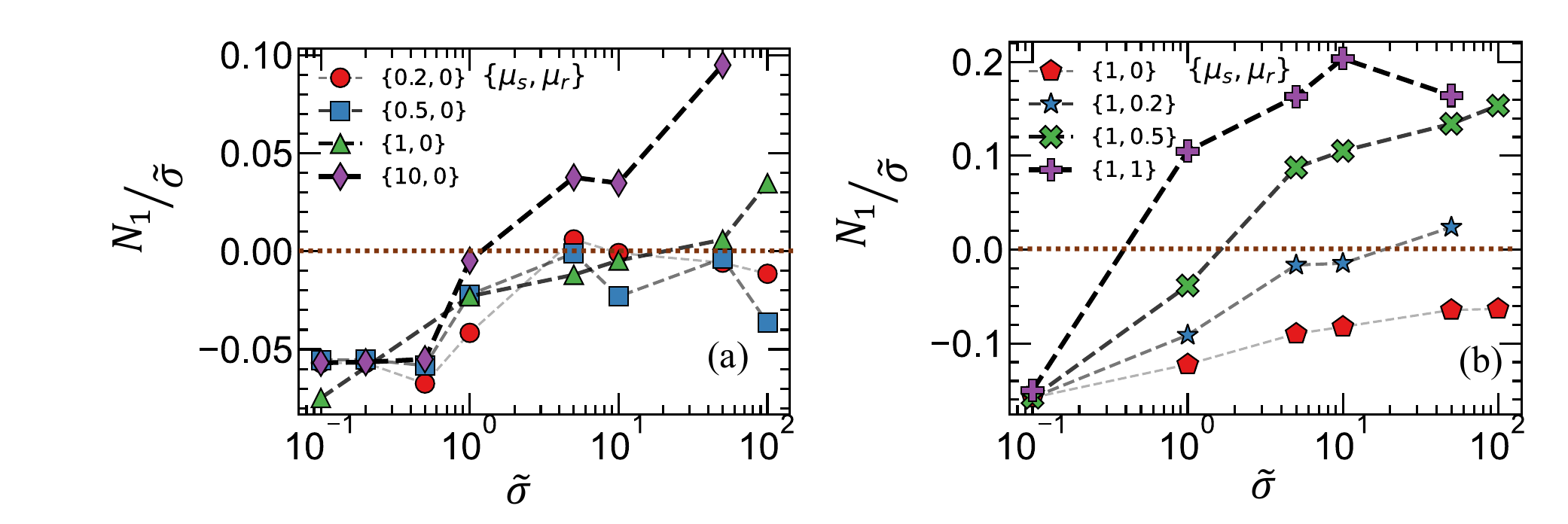}
    \caption{First normal stress difference coefficient $N_1$ normalized by shear stress $\tilde{\sigma}$ is plotted as a function for shear stress $\tilde{\sigma}$ for sliding constraint only at $\phi=0.78$ (a) and sliding and rolling constraints at $\phi=0.7$ for different rolling constraints (b). The dashed line is for $N1/\tilde{\sigma}=0$. At $\mu_s=10$, $N1/\tilde{\sigma}$ transitions to positive values, as depicted in (a). A similar positive shift is observed with increasing rolling friction, as shown in (b).}
    \label{fig:N1}
\end{figure*}
\clearpage
\subsection{Frictional contact network in dense suspensions with sliding-only constraints}

\begin{figure*}[h]
    \centering
\includegraphics[width=1\linewidth]{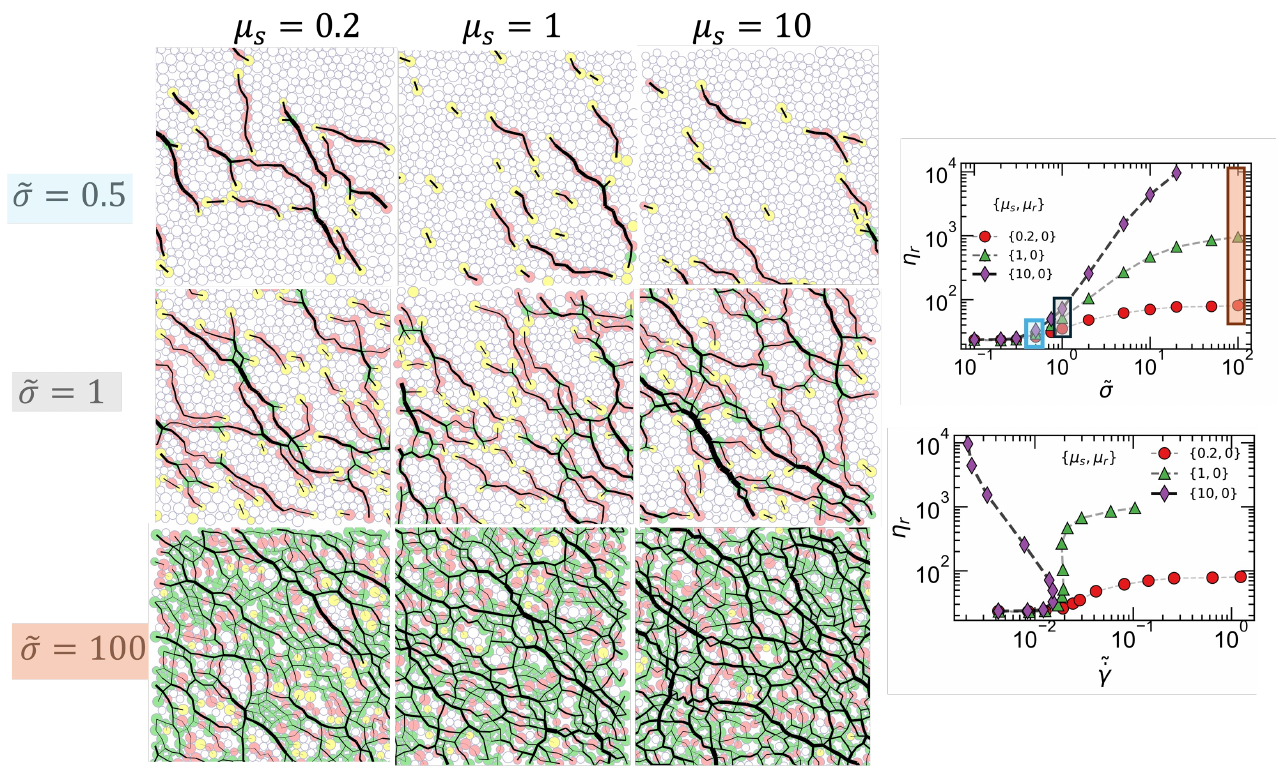}
    \caption{Microstructural snapshots of dense suspensions undergoing continuous and discontinuous shear thickening, as well as approaching shear jamming for sliding constraint only is shown for different sliding constraints $\mu_s=0.2$, 1 and 10 and $\mu_r=0$ and $\tilde{\sigma}=0.5$, 1, and 100. The corresponding flow curve is shown on the right of the figure. The microstructure snapshots illustrate the thickened states for CST ($\mu_s = 0.2$, $\tilde{\sigma} = 100$), DST ($\mu_s = 1$, $\tilde{\sigma} = 100$), and approaching shear jamming ($\mu_s = 10$, $\tilde{\sigma} = 100$). Particles are color-coded according to their local frictional coordination number ($Z_\mu^{\mathrm{local}}$), with white particles representing rattlers ($Z_\mu^{\mathrm{local}}$=0), yellow representing the terminals($Z_\mu^{\mathrm{local}}$=1), red particles representing spokes($Z_\mu^{\mathrm{local}}=2$), and green represent hubs($Z_\mu^{\mathrm{local}}\geq3$). Black lines denote frictional contacts that constitute force chains, where line thickness reflects the magnitude of the transmitted force.}
    \label{fig:snapshot_sliding}
\end{figure*}
\clearpage
\subsection{Frictional contact network in dense suspensions with sliding-and-rolling constraints}

\begin{figure*}[h]
    \centering
\includegraphics[width=1\linewidth]{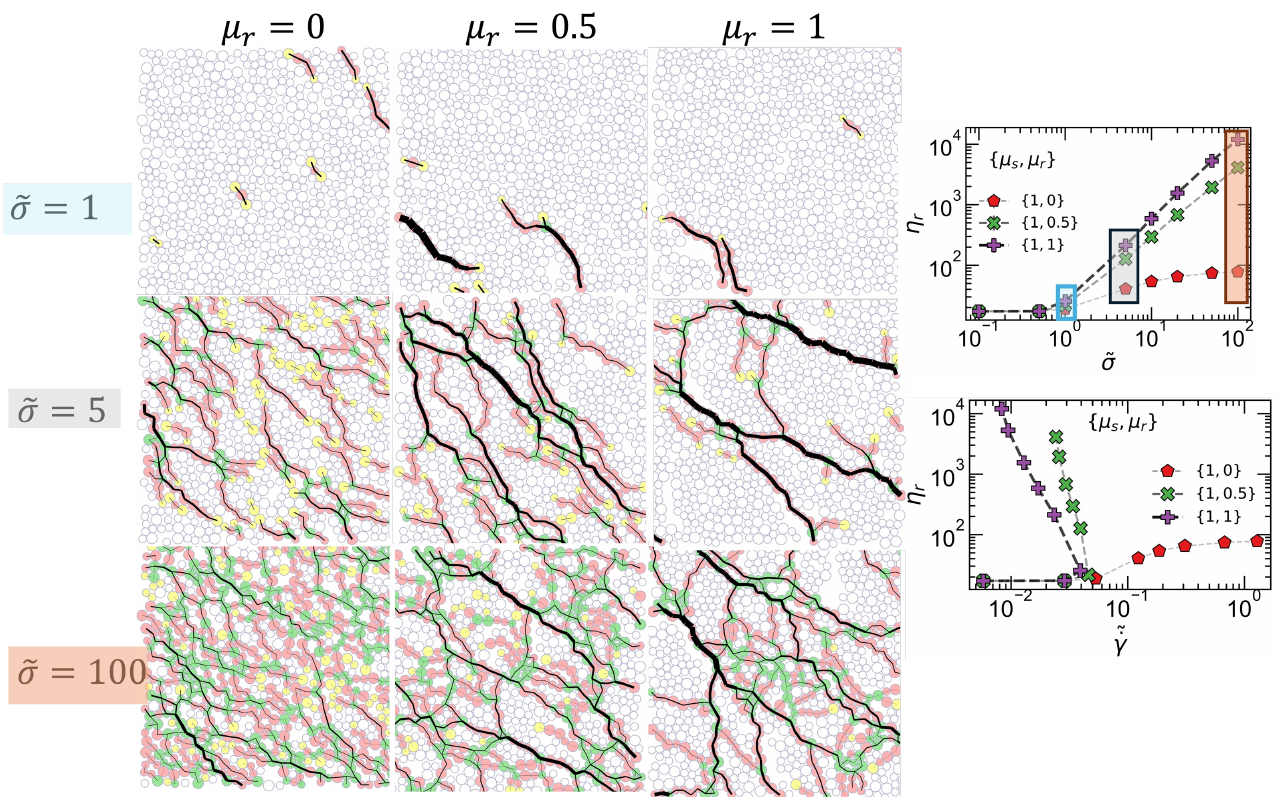}
    \caption{Microstructural snapshots of dense suspensions undergoing continuous and discontinuous shear thickening, as well as approaching shear jamming for sliding and rolling constraints is shown for different rolling constraints $\mu_r=0$, 0.5 and 1  with a fix sliding constraint $\mu_s=1$ and $\tilde{\sigma}=1$, 5, and 100 . The corresponding flow curve is shown on the right of the figure. The microstructure snapshots illustrate the thickened states for CST ($\mu_r = 0$, $\tilde{\sigma} = 100$), DST ($\mu_r = 0.5$, $\tilde{\sigma} = 100$), and approaching shear jamming ($\mu_r = 1$, $\tilde{\sigma} = 100$). Particles are color-coded according to their local frictional coordination number ($Z_\mu^{\mathrm{local}}$), with white particles representing rattlers ($Z_\mu^{\mathrm{local}}$=0), yellow representing the terminals($Z_\mu^{\mathrm{local}}$=1), red particles representing spokes($Z_\mu^{\mathrm{local}}=2$), and green represent hubs($Z_\mu^{\mathrm{local}}\geq3$). Black lines denote frictional contacts that constitute force chains, where line thickness reflects the magnitude of the transmitted force. }
    \label{fig:snapshot_rolling}
\end{figure*}

\clearpage
\section{Network Analysis}
\subsection{Fraction of Spokes}
\begin{figure*}[h]
    \centering
    \includegraphics[width=0.8\linewidth]{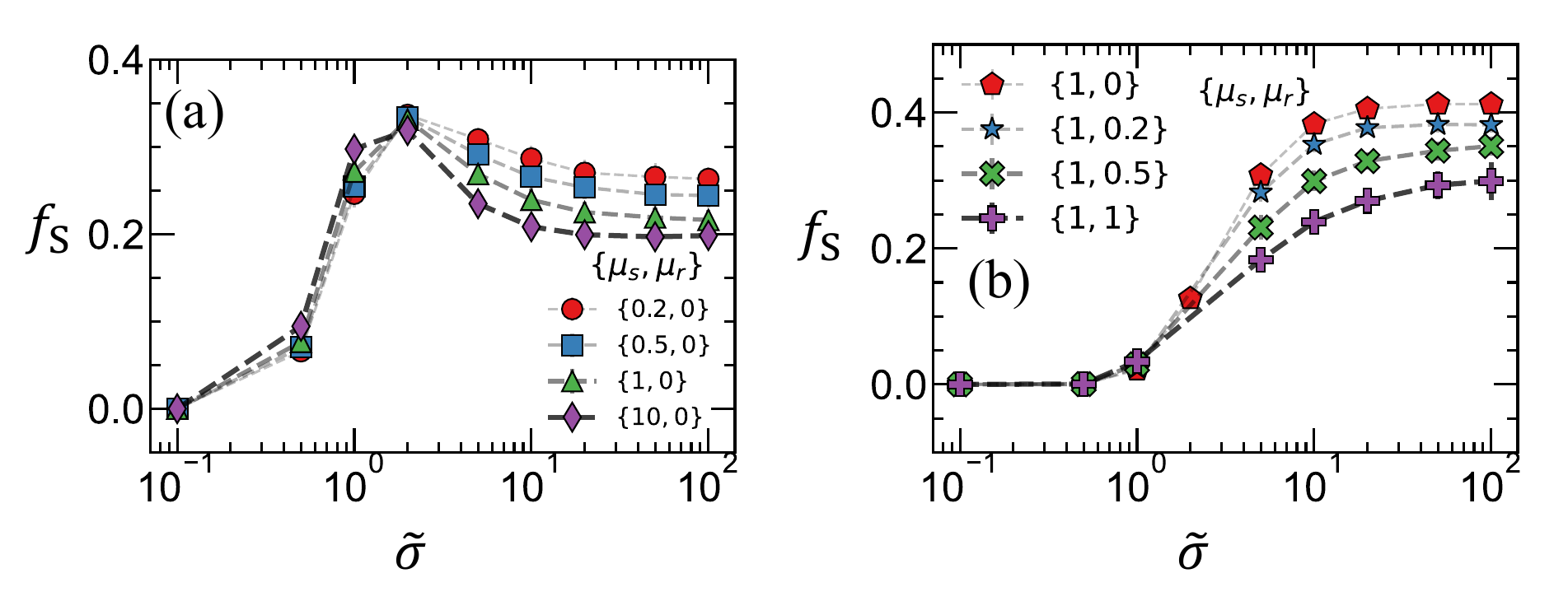}
    \caption{(a) Fraction of spokes $(f_\mathrm{S}-(Z^{\mathrm{local}}_{\mu}=2))$ is plotted as a function of stress $(\tilde
    {\sigma})$ for dense suspension with sliding constraint only at $\phi=0.78$ for different sliding constraints, $\mu_s$ ranging from 0.2 to 10. The flow curve, contact network, coordination number, fraction of hubs,terminals, and rattlers for this system are shown in Fig.\,3(a-f) and Fig.\,4(a-e). (b) Fraction of spokes $(f_\mathrm{S}-(Z^{\mathrm{local}}_{\mu}=2))$ is plotted as a function of stress $(\tilde
    {\sigma})$ for dense suspension with sliding and rolling constraints at $\phi=0.7$ for different rolling constraints, $\mu_r$ ranging from 0 to 1 and sliding constraint $\mu_s=1$. The flow curve, contact network, coordination number, fraction of hubs,terminals, and rattlers for this system are shown in Fig.\,3(g-l) and Fig.\,4(f-j). Fraction of spokes is of the similar order for both cases but it shows an overshoot for the sliding-only case while a gradual increase for sliding-and-rolling case. }
    \label{fig:spokes_sliding}
\end{figure*}
\clearpage

\subsection{Length of Edge or Force chains}
\begin{figure*}[h]
    \centering
    \includegraphics[width=0.8\linewidth]{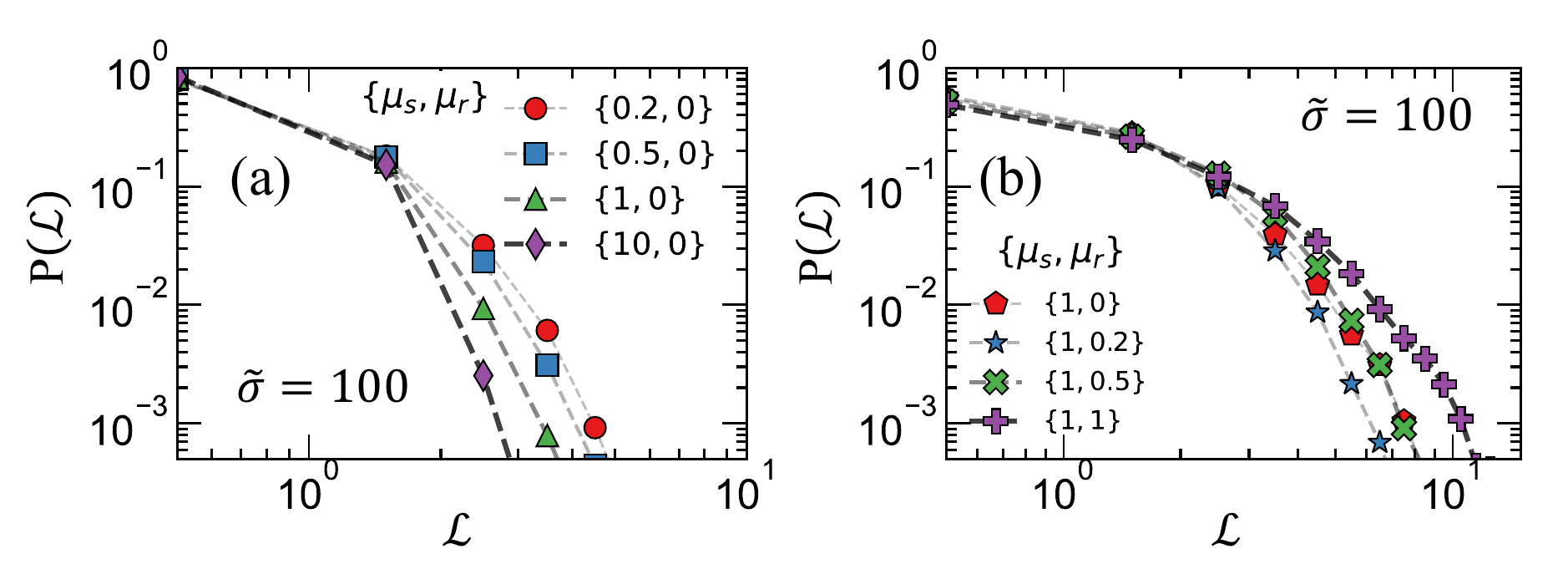}
    \caption{(a) Probability distribution of edge length at $\tilde{\sigma}=100$ for dense suspension with sliding constraint only at $\phi=0.78$ for different sliding constraints, $\mu_s$ ranging from 0.2 to 10. The edge length distribution for $\tilde{\sigma}=1-100$ for this system are shown Fig.\,4(d-e). (b) Probability distribution of edge length at $\tilde{\sigma}=100$ for dense suspension with sliding and rolling constraints at $\phi=0.7$ for different rolling constraints, $\mu_r$ ranging from 0 to 1 and sliding constraint $\mu_s=1$. The edge length distribution for $\tilde{\sigma}=1-100$ for this system are shown Fig.\,4(f-g). For both systems with similar bulk viscosity and approaching shear jamming, for the sliding-only case, the edge length get shorter with increase in $\mu_s$ but for the sliding-and-rolling case, edge length increases.}
    \label{fig:edge_length_sliding}
\end{figure*}

\clearpage

\section{Anisotropy of contact network}

\begin{figure*}[h]
    \centering
    \includegraphics[width=0.8\linewidth]{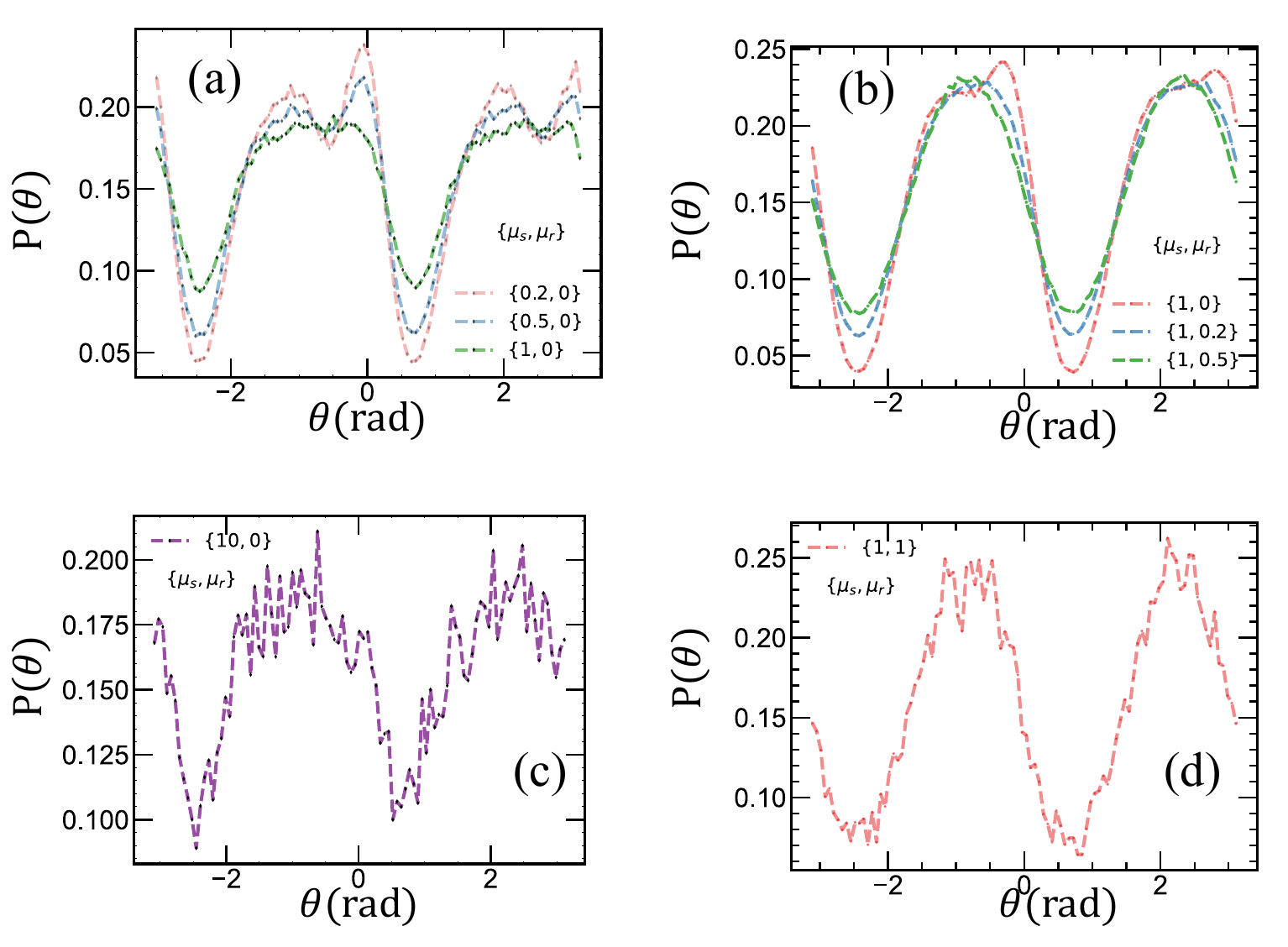}
    \caption{(a)\&(c)Probability distribution of particle orientation at $\tilde{\sigma}=100$ for dense suspension with sliding constraint only at $\phi=0.78$ for different sliding constraints, $\mu_s$ ranging from 0.2 to 10 (a-$\mu_s=0.2,0.5,1$) (c-$\mu_s=10$). The flow curve, contact network, coordination number, fraction of hubs,terminals, and rattlers for this system are shown in  Fig.\,3(a-f) and Fig.\,4(a-e). (b)\&(d) Probability distribution of particle orientation at $\tilde{\sigma}=100$ for dense suspension with sliding and rolling constraints at $\phi=0.7$ for different rolling constraints, $\mu_r$ ranging from 0 to 1 (b-$\mu_r=0,0.2,0.5$) (d-$\mu_r=1$). The flow curve, contact network, coordination number, fraction of hubs,terminals, and rattlers for this system are shown in  Fig.\,3(g-l) and Fig.\,4(f-j). For both systems, sliding only and sliding-and-rolling cases, the dense suspension shows anisotropic contact network as $P(\theta)$ is dependent on $\theta$ and this result is in agreement with studies reported in literature \cite{singh2017microstructural}. The variation of $P(\theta)$ versus $\theta$ also shows a secondary bump at low $\mu_s$ in the sliding case and at low $\mu_r$ in the sliding-and-rolling case and a similar feature has been observed in experimental studies \cite{lee2024particle}. }
    \label{fig:anisotropy_sliding}
\end{figure*}

\clearpage

\end{document}